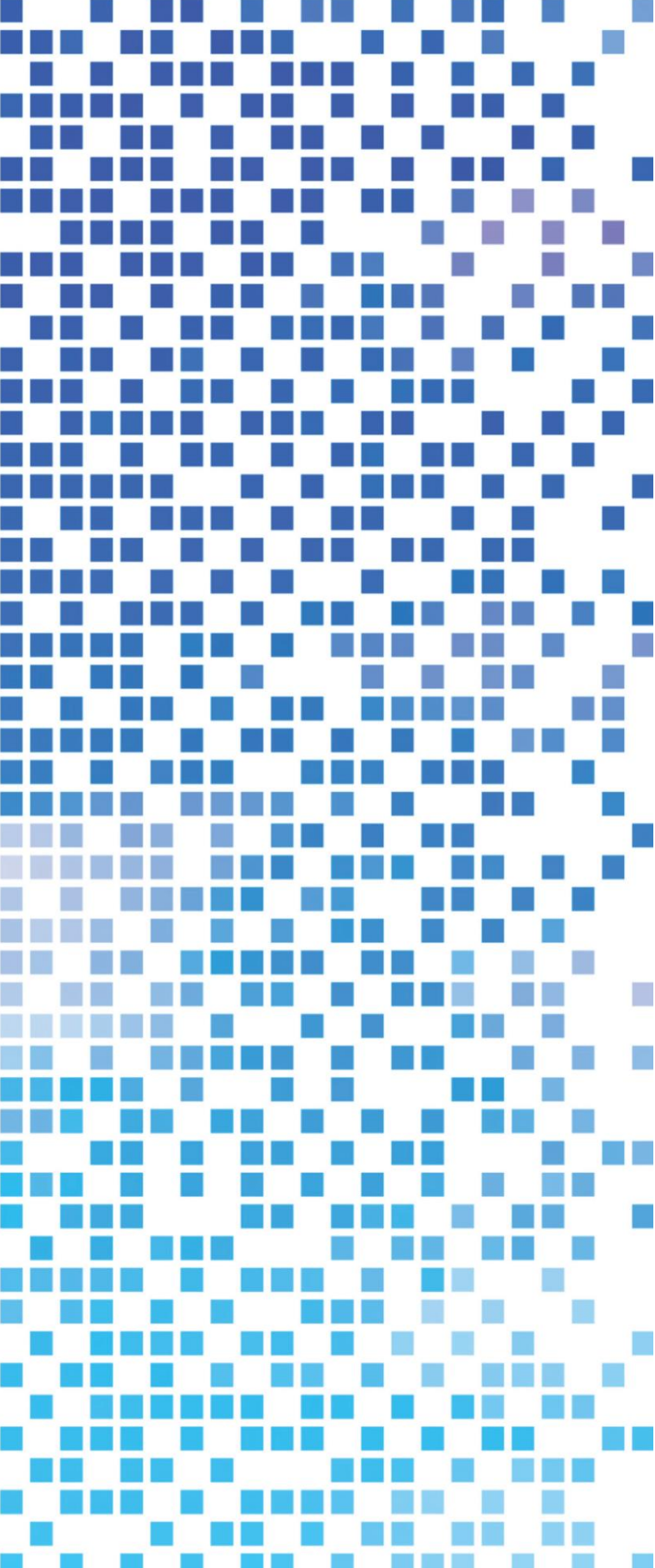

# ESCAPE

Preparing Forecasting Systems for the Next generation of Supercomputers

# D3.4 Performance report and optimized implementations of Weather & Climate dwarfs on multi-node systems

Dissemination Level: Public

This project has received funding from the European Union's Horizon 2020 research and innovation programme under grant agreement No 67162

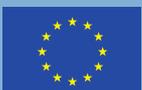
Funded by the European Union

Co-ordinated by

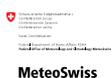 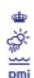 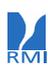 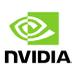 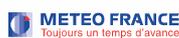 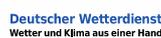 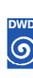 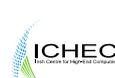 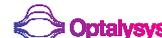 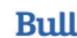 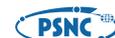 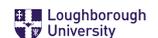

# ESCAPE

**Energy-efficient Scalable Algorithms for Weather Prediction at Exascale**

Author **Louis Douriez (Bull), Alan Gray (NVIDIA), David Guibert (Bull), Peter Messmer (NVIDIA), Erwan Raffin (Bull)**

Date **30/05/2018**

# Table of Contents

# Figures

# 1 Executive Summary

The goal of the ESCAPE project is to analyze and optimize time- and energy-to-solution for core components of numerical weather and climate simulation codes on modern hardware platforms. These components, so called *dwarfs*, are extracted from existing weather and climate codes, ported to accelerated systems by domain scientists and optimized by the hardware experts so that they can ultimately be used as building blocks or design guidance for next generation numerical weather and climate code.

Here we summarize the work performed on optimizations of the dwarfs focusing on CPU multi-nodes and multi-GPUs. We limit ourselves to a subset of the dwarf configurations chosen by the consortium.

Intra-node optimizations of the dwarfs and energy-specific optimizations have been described in Deliverable D3.3.

To cover the important algorithmic motifs we picked dwarfs related to the dynamical core as well as column physics. Specifically, we focused on the formulation relevant to spectral codes like ECMWF's IFS code.

The main findings of this report are:

- Up-to 30% performance gain with CPU based multi-node systems compared to optimized version of dwarfs from task 3.3 (see D3.3).

- Up to 10X performance gain on multiple GPUs from optimizations to keep data resident on the GPU and enable fast inter-GPU communication mechanisms.

- Multi-GPU systems which feature a high-bandwidth all-to-all interconnect topology with NVLink/NVSwitch hardware are particularly well suited to the algorithms.

While a direct comparison of CPU and GPU accelerated implementations for some dwarfs is desirable, doing a realistic comparison is very difficult as a lot of aspects have to be taken into account. This report therefore proposes complementary approaches.

Both CPU and GPU approaches used in this work are complementary as a hybrid system will take advantage of most of optimizations presented in this report.

# 2 Introduction

## 2.1 Background

ESCAPE stands for Energy-efficient Scalable Algorithms for Weather Prediction at Exascale. The project develops world-class, extreme-scale computing capabilities for European operational numerical weather prediction and future climate models. ESCAPE addresses the ETP4HPC Strategic Research Agenda 'Energy and resiliency' priority topic, promoting a holistic understanding of energy-efficiency for extreme-scale applications using heterogeneous architectures, accelerators and special compute units by:

- Defining and encapsulating the fundamental algorithmic building blocks underlying weather and climate computing;

- Combining cutting-edge research on algorithm development for use in extreme-scale, high-performance computing applications, minimizing time- and cost-to-solution;
- Synthesizing the complementary skills of leading weather forecasting consortia, university research, high-performance computing centers, and innovative hardware companies.

ESCAPE is funded by the European Commission's Horizon 2020 funding framework under the Future and Emerging Technologies - High-Performance Computing call for research and innovation actions issued in 2014.

The objectives of ESCAPE are met by a cyclic collaboration of individual work package teams (WP): WP1 focuses on the extraction and definition of the dwarfs, WP2 on porting to accelerated architectures, WP3 on the optimization for individual target platforms. The insights and results gained from this work are then fed back into WP1 for a refined definition of the dwarfs.

In this report we present the work performed by WP3 on the port and optimization of selected dwarfs to multi-node CPU based systems and multi-GPU ones.

This work directly contributes to achieving the project's top-level objective 2, diagnose and classify weather & climate dwarfs on different HPC architectures, and 3, combine frontier research on algorithm development and extreme-scale, high-performance computing applications with novel hardware technology.

## 2.2 Scope of this deliverable

### 2.2.1 Objectives of this deliverable

The objective of this deliverable is to report the different optimizations done in ESCAPE on selected dwarfs and the corresponding performance results on state of the art multi-node CPU and multi-GPU systems.

### 2.2.2 Work performed in this deliverable

The work performed for this deliverable consisted mostly of task 3.4 of the ESCAPE Description of Action (DoA). This includes optimizations of the dwarfs on multi-node CPU and multi-GPU systems.

A selection of dwarfs has been defined according to their relevance for the NWP centers, especially the spectral transform dwarfs which represents the most important challenge in the Integrated Forecasting System (IFS) of ECMWF as reported in D1.7 ("Assessment report on the efficiency of a multi-resolution, spectral transform approach to future global modelling").

Starting from the optimized version from task 3.3, different optimization tracks targeting multi-device systems have been used to improve overall performance.

In addition to arriving at an optimized implementation of specific dwarfs, the project aims at determining best practices for implementing numerical methods on different architectures.

### 2.2.3 Deviations and counter measures

The targeted systems for this work were initially Nvidia GPUs and Intel Xeon Phi accelerators. During the course of the project, the Xeon Phi product line dedicated to the HPC market was discontinued. We therefore decided to concentrate our efforts on modern Xeon based multi-node systems. This choice still allows tackling the main addressed challenge: multi-node performance and especially communication aspects of NWP skeleton apps on state-of-the-art HPC systems.

#### 2.2.4 Organization of the report

The rest of the report is organized as follows: for each selected dwarf we present the optimization strategy for the targeted system followed by the optimization results and analysis, and summarize the findings and further optimizations in a concluding section.

## 3 Spectral Transform – Spherical Harmonics Dwarf

### 3.1 Multi-GPU Optimization

#### 3.1.1 Introduction

In deliverable D3.3 we presented single GPU optimizations of the Spherical Harmonics Dwarf on the NVIDIA GPU architecture. In this deliverable we extend these optimizations to the multi-GPU case. One of the critical parts in scaling the Spherical Harmonics dwarf to multiple GPUs is that at every timestep, each GPU sends and receives data from every other GPU. This all-to-all communication requires a high injection bandwidth network between the GPUs to maintain scalability of the code. We will show that the modern NVLink interconnect hardware, in combination with the recently announced NVSwitch, offers the necessary injection bandwidth for this operation, enabling efficient multi-GPU scaling. However, to leverage these novel networking hardware components it is important to use a highly optimized code. We therefore describe the multi-GPU performance optimizations for the spherical harmonics dwarf in detail below and describe the impact on the overall performance.

In the Architecture section we give details of the DGX-1 and DGX-2 servers used in these investigations. In the Multi-GPU Optimization and Results section we describe the optimizations and present results showing the effect of these. Finally, in the Power and Energy Utilization section, we present a short analysis of power and energy utilization of the optimized code on multiple GPUs.

#### 3.1.2 Architecture

When running the Spherical Harmonics Dwarf across multiple GPUs, all-to-all communications are required at every timestep. The performance of these operations are limited by the lowest-bandwidth link in use, so for optimal performance we need a network topology whereby each GPU is connected to each other GPU with maximum bandwidth. Each GPU features multiple ports of high-bandwidth NVlink connections, each providing 50 GB/s of bi-directional bandwidth (when using the Volta GPUs; or 40 GB/s for the older Pascal generation).

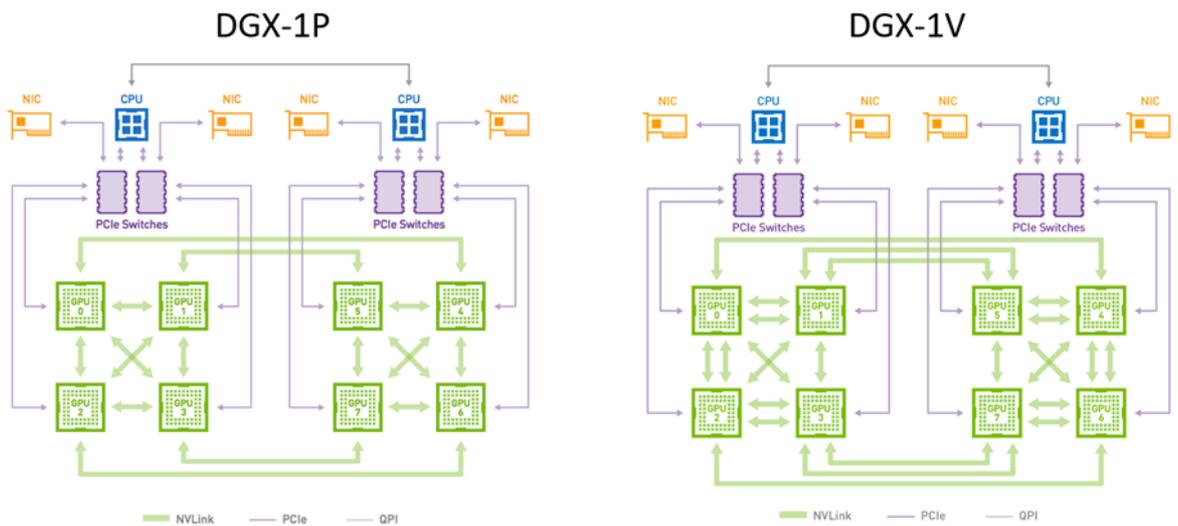

*Figure 1: The NVLink architecture of the DGX-1 server, with the Pascal and Volta variants shown on the left and right respectively.*

In Figure 1 we show the architecture of the DGX-1 server. The main point relevant here is that all-to-all connectivity only exists on each 4-GPU "island" in the server, not across all 8 GPUs (e.g GPUs 0 and 5 do not have a direct link). (Note that the "extra" NVLinks on the Volta variant, arising from the extra ports on the V100 chip, are not relevant for this study since the minimum number of GPUs in use is 4 due to memory requirements; the performance of all-to-all is dictated by the lowest-bandwidth connection; and we don't have full double-link-connectivity even on each 4 GPU island.) For full bandwidth connectivity when using more than 4 GPUs, extra interconnect hardware is required: the newly announced NVSwitch interconnect as featured on the DGX-2 server fulfils this purpose.

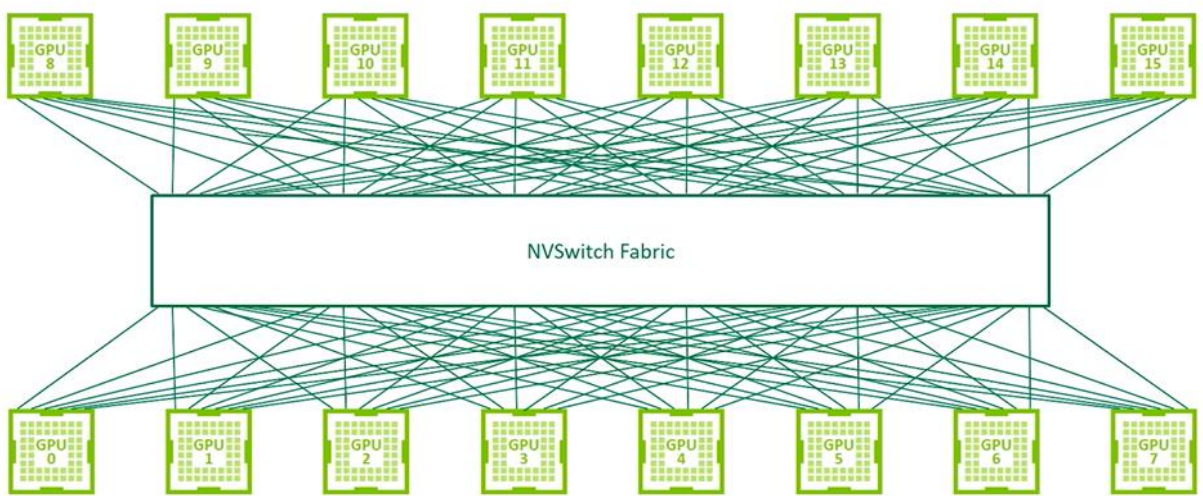

*Figure 2: The NVlink/NVSwitch architecture of the DGX-2 server.*

In Figure 2 we show how NVSwitch allows all-to-all connectivity to extend to many more GPUs than previously possible. The DGX-2 server has 16 Volta V100 GPUs: each with six 50 GB/s NVLink connections into the switch with routing to any of the other GPUs in the system. This allows 300 GB/s communications between any pair of GPUs in the system, or equivalently 2.4 TB/s total throughput.

### 3.1.3 Optimization strategies

#### 3.1.3.1 CUDA-aware MPI

When running a single application across multiple GPUs, it is necessary to transfer data between the distinct memory spaces. Traditionally, such transfers needed to be realized via host memory and required the participation of the host CPU. Not only did this introduce additional latency, but also limited the overall bandwidth to the bandwidth offered by the PCIe bus connecting CPU and GPUs. However, modern MPI implementations are CUDA-aware. This means that pointers to GPU memory can be passed directly into the MPI calls, allowing the avoidance of unnecessary transfers (both in the application and in the underlying MPI implementation). This is particularly useful when using a server that features high-bandwidth NVLink connections between GPUs, in which case CUDA-aware MPI will use these links automatically.

Using CUDA-aware MPI is relatively simple. Consider a non-CUDA aware code that performs some MPI communication, e.g. MPI_Alltoallv:

```
!$ACC update host(array,…)  ! copy array from GPU to CPU
call MPI_Alltoallv(array,…) ! CPU-side MPI communication
!$ACC update device(array,…). ! copy array from CPU to GPU
```

This can be adapted to use CUDA-aware MPI as follows:

```
!$ACC host_data use_device(array,…) ! Instruct OpenACC to use GPU pointer directly
call MPI_Alltoallv(array,…) ! GPU-side MPI communication
!$ACC end host_data
```

However, the situation can be more complicated (as is the case with the Spherical Harmonics dwarf) if explicit buffer packing is involved within the application before and after communications, which then also needs to be ported from CPU to GPU using OpenACC in the usual manner to allow data to be kept resident on the GPU throughout.

#### 3.1.3.2 Optimization of all-to-all communications

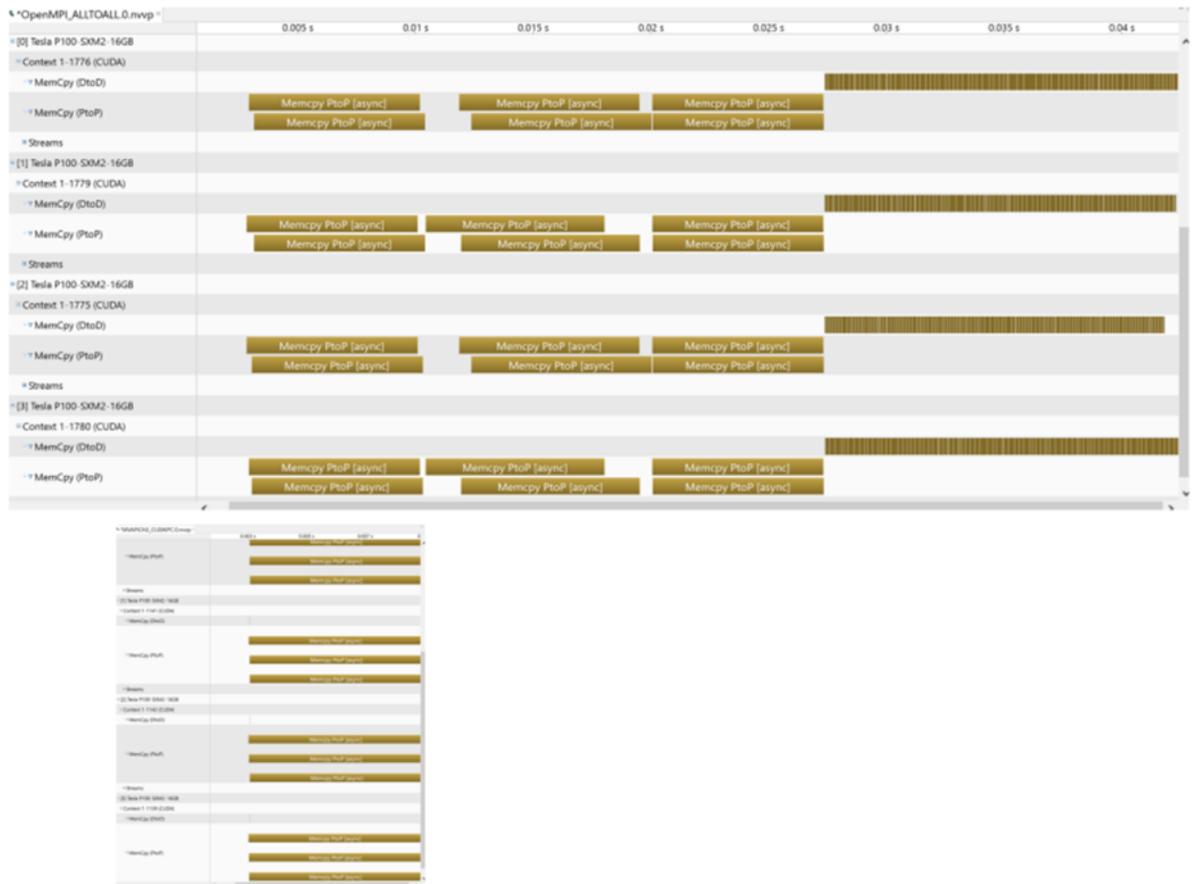

*Figure 3: The nvprof timeline for all-to-all communication using CUDA-aware MPI (top) and a custom implementation using CUDA IPC and streams (bottom). The top and bottom figures have been lined up such that the timelines match, to highlight the advantage of the custom implementation.*

The first multi-GPU optimization of the Spherical Harmonics Dwarf was the introduction of CUDA-aware MPI. However, even with this optimization the all-to-all operations remained inefficient. Figure 3 (top) shows a profile of the all-to-all operation on 4 GPUs within a DGX-1P server. Inspecting the timeline from left to right, the first three stages correspond to each GPU performing a memory copy to each other GPU via NVLINK. The final fourth stage corresponds to the transfer to self. This pattern is suboptimal, as the system exhibits separate NVLink connections between each pair of GPUs within each 4 GPU partition of the system, allowing all these messages to be exchanged concurrently.

We therefore implemented an optimized version of the all-to-all communication phase directly in CUDA using the Inter Process Communication (IPC) API. Using memory handles, rather than pointers, CUDA IPC allows to share memory spaces between multiple processes, thus allowing one GPU to directly access memory on another GPU. Together with the CUDA streams facility which allows overlapping of independent operations, the desired overlap was achieved. The steps are as follows:

Setup:

- On each MPI rank, get IPC memory handle for array locally.

- Share IPC memory handles via MPI.
- Setup CUDA streams.

all-to-all:

- On each rank, loop over all `targetranks` (including own).
- `targetrank=(targetrank+rank)%number_of_gpus` (for better balance).
- Push message to `targetrank (in stream[targetrank])`.
- Synchronize all streams.

This allows the desired overlapping to occur, as can be seen in the bottom part of Figure 3 which has been lined up such that the timeline matches the original case to emphasize the performance improvement achieved for this communication phase. This improvement will be put into context within the full application timestep in the next section.

### 3.1.4 Results

The results presented in this section have been obtained using the following software configuration: PGI 18.1, CUDA 9.1.85 and OpenMPI 3.1.0.

#### 3.1.4.1 Results using 4 GPUs on DGX-1V

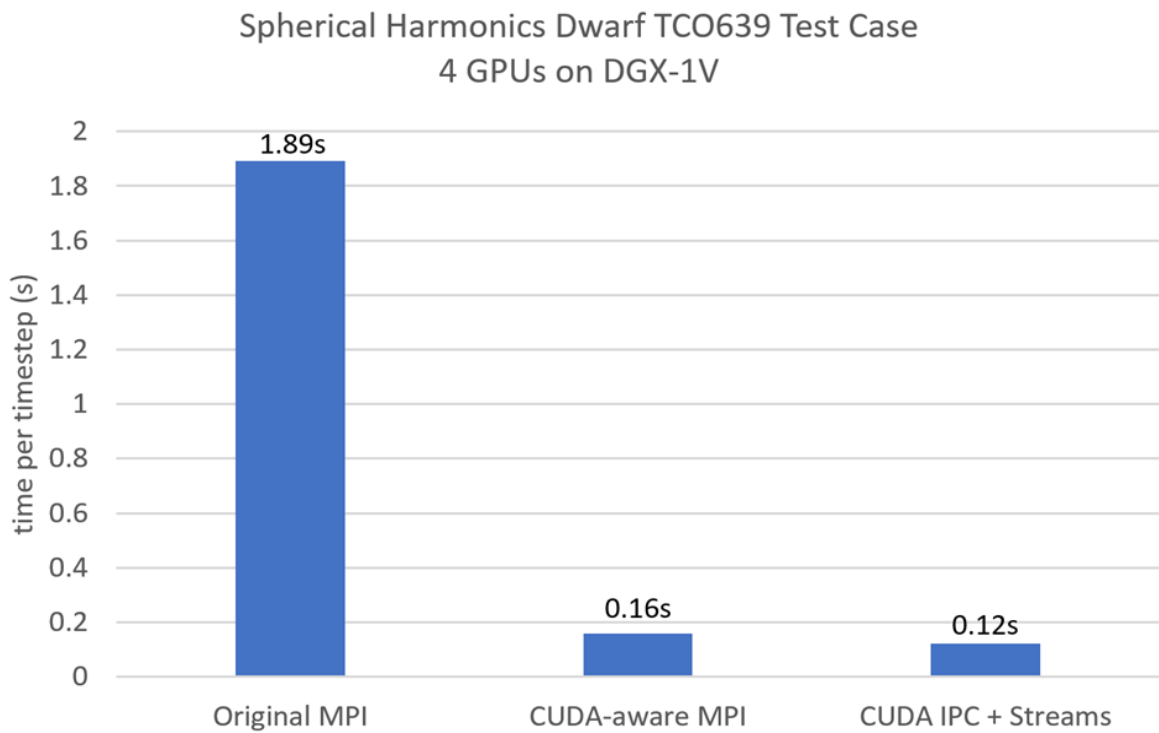

*Figure 4: The dependence of the Spherical Harmonics TCO639 test case time on the all-to-all communication method used. Compared are, from left to right, use of the "original" code which used non CUDA-aware MPI and buffer operations on the CPU, CUDA-aware MPI with buffer operations on the GPU, and a custom GPU implementation using a combination of CUDA IPC and CUDA streams.*

In Figure 4 we show the results of communication optimization for the TCO639 test case on 4 GPUs (the minimum number required to provide the necessary memory), on the DGX-1V server. The leftmost column shows the wallclock time per timestep when using the original communication mechanisms in the code (where the computational parts of the code are optimized, as described in the previous deliverable D3.3). The use of CUDA-aware MPI together with on-device buffer packing and unpacking (center column) is seen to provide a dramatic improvement in performance: from 1.89s to 0.16s overall. A further boost is provided by the replacement of MPI_Alltoallv with our new custom all-to-all code written using CUDA IPC and CUDA streams, which brings the time down to 0.12s.

The above results use the fully-NVLink-connected 4 GPU partition of DGX-1V. However, when we want to scale beyond 4 GPUs on the DGX-1 platform a problem arises. It can be seen from Figure 1 that, when using all 8 GPU in DGX-1V, we no longer have full NVLink connectivity. For example, there is no direct link between GPU 0 and GPU 5. So, for an all-to-all operation, some messages travel through the relatively low bandwidth PCIe bus (connecting GPU to CPU) and QPI link (connecting the two CPUs together): this limits performance. Furthermore, when using multiple DGX-1V servers (e.g. 16 GPUs across 2 servers), some messages must go across Infiniband network, which is also has low bandwidth relative to NVLink.

### 3.1.4.2 Multi-GPU scaling with NVSwitch on DGX-2

To address the fact that an increasing number of problems are requiring an increasing number of tightly-connected GPUs, at the GTC 2018 conference NVIDIA announced a new interconnect: NVSwitch. This allows many more GPUs to communicate in a symmetric fashion at higher bandwidth than previously possible. Also announced was the 16-GPU DGX-2 server that incorporates NVSwitch to allow full-NVLINK-bandwidth communications between all GPUs in the system, as shown in Figure 2. This provides an ideal architecture for the Spherical Harmonics dwarf, since all bottlenecks have been removed from the all-to-all communications, allowing the code to scale well and efficiently use all of the available GPUs in parallel.

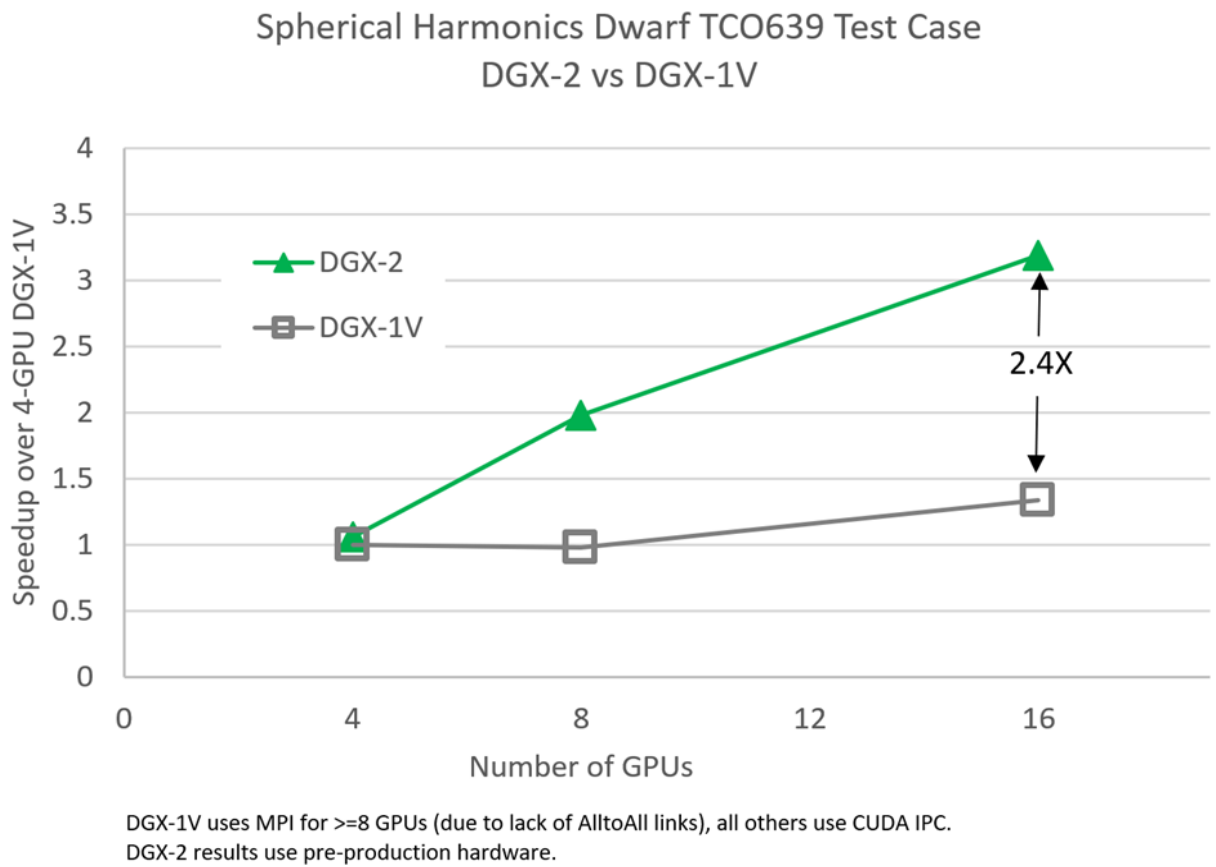

DGX-1V uses MPI for >=8 GPUs (due to lack of AlltoAll links), all others use CUDA IPC.
DGX-2 results use pre-production hardware.

*Figure 5: The dependence of the speedup of the Spherical Harmonics TCO639 test case on the number of GPUs in use. The grey squares denote results obtained using DGX-1V hardware whilst green triangles denote DGX-2.*

In Figure 5 we demonstrate how the use of DGX-2 with NVSwitch allows significantly better scaling that use of DGX-1 for the Spherical Harmonics TC0639 test case. Note that we tune the number of MPI tasks in use: we use the NVIDIA Multi Process Service to allow oversubscription of GPUs such that, e.g. the 8 GPU result on DGX-2 uses 16 MPI tasks across the 8 GPUs (i.e. 2 operating per GPU). This is because such oversubscription can sometimes be beneficial to spread out any load imbalance resulting from the spherical grid decomposition (see below) and hide latencies. We chose the best performing number of MPI tasks per GPU in each case.

It can be seen that, using only 4 GPUs, the results across these systems are very similar: this is because in both cases we have full NVLink connectivity between each of the four GPUs in use. However, as we increase the number of GPUs, the scaling on DGX-1V is limited: This is because we no longer retain full connectivity and some messages must go through the lower-bandwidth PCIe and QPI links and/or Infiniband when scaling across multiple servers. But on DGX-2 with NVSwitch, all 16 GPUs have full connectivity: that is we have maximum peak bandwidth of 300 GB/s between each pair of GPUs in use. The performance is seen to scale well out to the full 16 GPUs on DGX-2, where the difference with the 16 GPU (2-server) DGX-1V result is 2.4X.

It can also be seen that the speedup going from 4 to 16 GPUs on DGX-2 is 3.2X, whereas the ideal speedup would be 4X. However, initial investigations reveal that this deviation from ideal scaling is not primarily due to communication overhead but instead to load imbalance between the MPI tasks from the spherical grid decomposition that is chosen by the application in each case, which would indicate that better scaling would be observed with a more balanced decomposition. This warrants more in-depth investigation once the DGX-2 server becomes fully-available.

### 3.1.4.3 Power and energy utilization

In this section we address the issue of power utilization, and in-turn energy usage, through measurements taken whilst running the code.

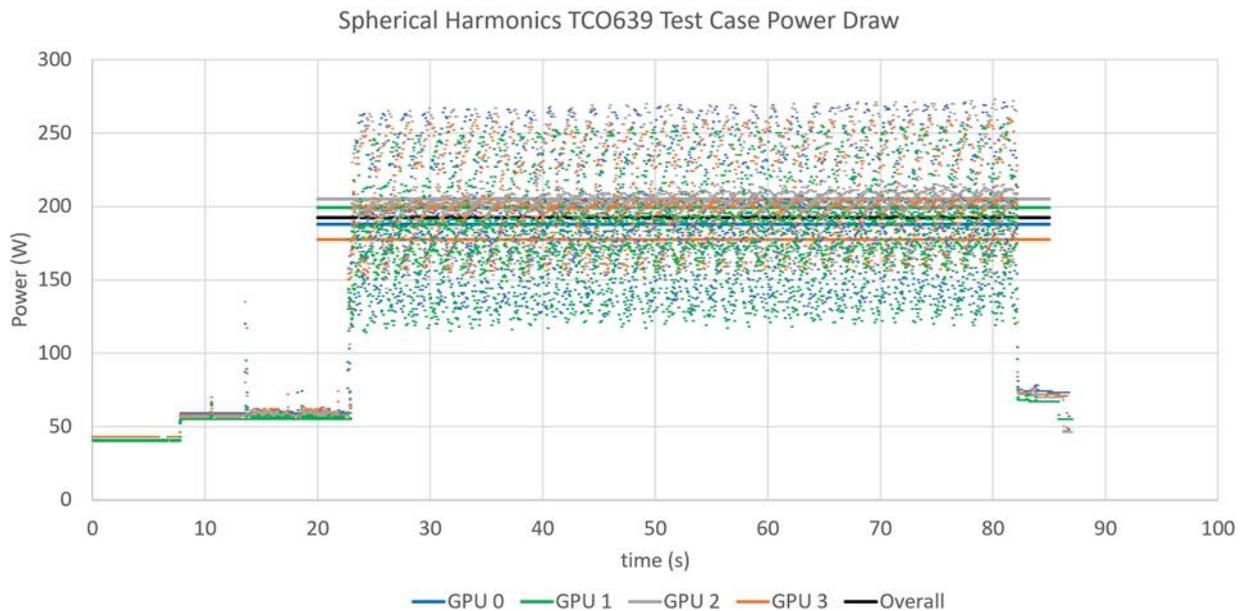

*Figure 6: The Power drawn by the Spherical Harmonics test case, as determined through use of the nvidia-smi monitoring tool. The four colours (blue, green, grey and orange) correspond to the 4 GPUs used to run the code: dots are samples and lines are averages. The black line is the average across all GPUs.*

In Figure 6 we show power measurements taken whilst running the code using the nvidia-smi tool. We used 4 V100 GPUs (within a DGX-1 server) and ran 500 timesteps, which correspond to around 1 minute of runtime: this region can clearly be seen on the plot (with initialization and finalization occurring before and after this region respectively). The maximum power rating of the V100 is 300W, and the samples taken within the time-stepping region are seen to vary in the range of 110-270W. There is a significant difference across GPUs as is seen by the averages (shown by the solid lines) of 188W, 200W, 205W and 178W for GPUs 0, 1, 2 and 3 respectively. These differences are due to load imbalance resulting from the spherical grid decomposition: the higher the average the less time spent in synchronization barriers. The overall average is shown by the black line at 193W.

Therefore, we can obtain estimate the energy usage of the dwarf by multiplying this average power draw (193W) by the time taken per timestep (0.12s), by the number of GPUs (4) as 93J per timestep. In general, by far the most effective way to reduce energy usage is to reduce the time taken by the simulation through performance optimization, as we have done to obtain this figure.

This approach obviously has the caveat that we are only including the power consumed by the GPUs and not by the other resources on the node (most notably CPUs), and a more accurate analysis would require node-level measurements. However, there is a trend for nodes to become "fatter" in terms of GPU to CPU ratios (e.g. the DGX-2 has 16 GPUs and only 2 CPUs), and the CPUs are mostly idle since we have ported all key kernels to the GPU. We believe that the GPUs are the main source of power draw on the node, so the above number gives a reasonable indication.

### 3.1.5 Summary

When utilizing multiple GPUs in parallel, the Spherical Harmonics Dwarf requires all-to-all data movement such that every GPU must send data to and receive data from every other GPU in use. For good performance this requires that the interconnect topology is such that all-to-all high-bandwidth connections are available in such an all-to-all. On a DGX-1 server, the dwarf performs well using a fully-connected 4-GPU island but scaling is limited beyond this scale. However, the newly announced DGX-2 server with NVSwitch all-to-all connectivity allows the dwarf to scale to all 16 GPUs.

In order to avoid inefficiencies at the application level, several optimizations were required and these were described with the associated performance improvements presented: the application was adapted to use CUDA-aware MPI (with all buffer packing and unpacking ported to the device), and a new custom implementation of all-to-all using CUDA IPC and CUDA streams was developed. We also presented results obtained through power measurements, to get an indication of energy utilization.

The results presented above give indications that multi-GPU systems can be effectively exploited for spherical harmonic simulations.

## 3.2 CPU based Multi-node Optimization

In this section, each CPU-based node used contains a bi-socket Intel® Xeon® Gold 6148 CPU at 2.40GHz with 192GB of memory per node at 2666MHz. Nodes are interconnected with Mellanox EDR hardware in a fat-tree topology.

MPI implementation, Fortran and C/C++ compilers and BLAS/LAPACK routines are from Intel 18.0.2 libraries.

System optimizations from D3.3, transparent huge page, turbo, and mmap threshold are used.

### 3.2.1 Optimization strategies

The previous deliverable D3.3 was focused on intra-node optimizations notably with Intel KNL and KNC. Despite KNL being discontinued, its AVX-512 instruction set is now available with the latest generation of Intel processors (Skylake). Previous work of non-intrusive vectorizations based on directives allows us to reuse KNL optimizations for Skylake without any code modification and stay compatible with non-AVX512 processors.

Inter-node performance of this dwarf is dominated by communication as visible on the following figure.

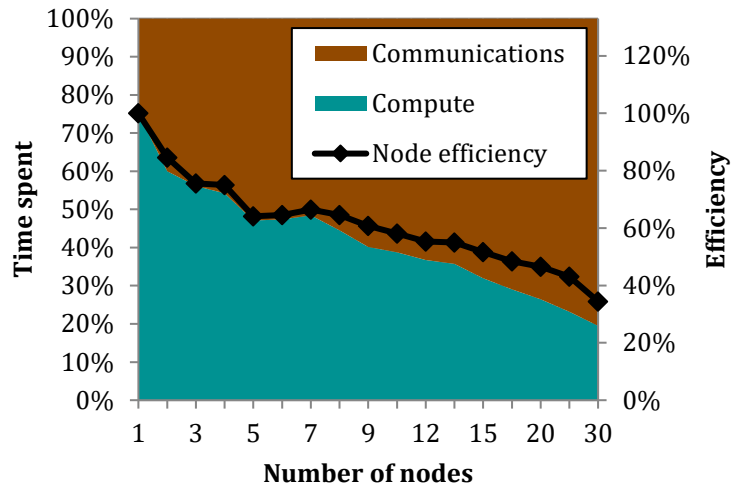

*Figure 7 : Global time repartition for TCo639 running on multiple bi-sockets node (Intel(R) Xeon(R) Gold 6148) on Hybrid MPI/OpenMP version.*

Communication can be split into two groups, point-to-point and collectives. In this dwarf, point-to-point communication can be again divided into data preparation and actual transfer. This separation is often fuzzy but the following figure (Figure 8) on the left shows that preparation was taking most of the communication time for small node counts.

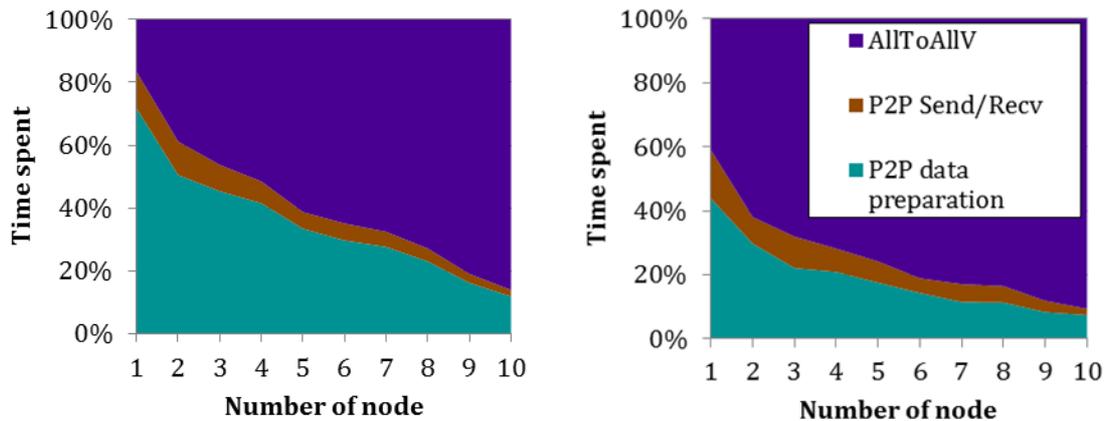

*Figure 8 : Communications time repartition for TCo639 running on multiple bi-sockets nodes (Intel(R) Xeon(R) Gold 6148), without optimization (left) and with optimizations (right).*

### 3.2.1.1 Runtime level optimizations

The base version used for the Spectral Transform - Spherical Harmonics dwarf come from the intra-node deliverable D3.3 where we demonstrated that on a bi-socket Skylake node, the version using only OpenMP directives obtained the best performance. However, with multiple interconnected nodes, the same strategy leads to bad efficiencies and even a slow-down. The domain decomposition through MPI processes leads to a large imbalance. Binding one node per MPI process is too coarse a granularity, especially if there are only a small number of MPI processes.

It leads us to choose a finer MPI granularity for inter-nodes runs. For a small number of nodes, the best performance was achieved using MPI only with one physical core per MPI rank. State-of-the-art MPI implementations, notably IntelMPI are sufficiently smart to detect and transfer data with a low overhead between shared memory MPI processes. This has been made possible via copy buffers for small messages (low latency copy-in/copy-out) and kernel zero-copy mechanisms for larger messages [1][2] or even direct cross mapping of physical pages between MPI processes [3]. The shared memory MPI module renders this mostly transparent to the user without any code modification even if the MPI standard 3.0 introduced some explicit syntax with one-sided communications. By disabling OpenMP we also remove a lot of synchronization due to its inherent fork-and-join model. Moreover, cache-level false sharing and NUMA first touch memory effects are both removed. However, for a larger number of nodes the hybrid MPI/OpenMP version allows to decrease global communication impact and memory usage and still improve computational performance.

For the hybrid version and for each MPI process, its OpenMP fork domain is bound to a level 2 cache. It allows two OpenMP process per threads and could lead to false and true sharing impact on L2 cache. Above two, we bind to a socket domain and therefore limit NUMA effect.

### 3.2.1.2 Point-to-Point communications

At the software level there are 3 pathways of optimizations for point-to-point communications, namely:

- Preparation algorithm (mostly Pack/Unpack)
- Communication algorithm (rcache, pre-registered buffers, pipelining, kernel zero-copy mechanisms, page crossmapping, )
- Overlapping communication notably with preparation

During the preparation phase of a point-to-point communication, the sender side gathers the local data into a contiguous buffer (Pack operation) and hands it off to the MPI library. On the receiver side, data is then scattered from a contiguous user buffer to its correct location (Unpack operation). Pack and Unpack are nearly inevitable with scattered data because Remote Direct Memory Access (RDMA) with no gather scatter operations are known to be often less effective notably due to the memory pinning latency [4]. It also means that sender and receiver must share their memory layout as they may differ. A concrete implementation of this last point is presented later in this deliverable with GPU optimization for MPDATA but limited to a shared memory context. For this dwarf and for CPU, the Pack and Unpack algorithms were scanning memory multiple times. We reduced it with a global performance gain on the whole application varying between 10% and 20% (TCo639).

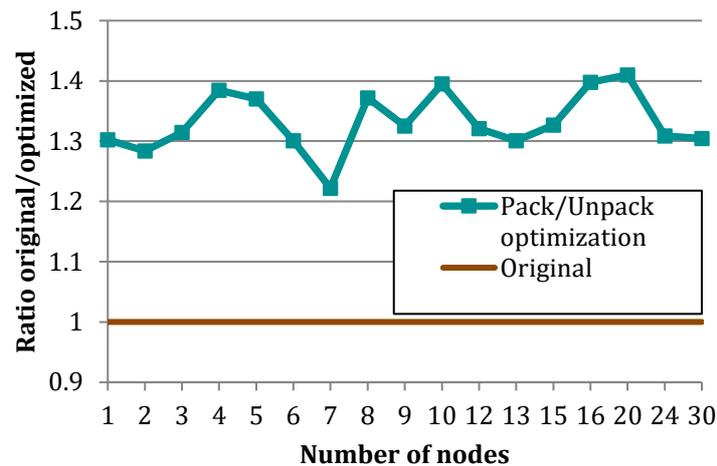

*Figure 9 : Performance improvement with new Pack/Unpack (TCo639)*

Communication algorithms rely on the latest MPI implementation from Intel. The runtime switches flawlessly between shared memory and network exchanges (through Mellanox EDR interconnect in our case) for respectively intra-nodes and inter-nodes point-to-point communications without code modification.

For inter-nodes exchanges, overlapping is already done by the IntelMPI library which pipelines exchanges and copy of user contiguous buffers into pre-registered/pre-pinned buffers. This extra copy, with the assertion that memory bandwidth is higher than the network bandwidth, is comparable with RDMA on-the-fly memory pinning in terms of performance and does not require to pre-allocate a send and receive buffer [5].

Overlapping between Pack/Unpack and data exchange can be implicit via MPI syntax but it doesn't guarantee any performance gains due to an increase in memory usage and a non-negligible overhead. Although this could be investigated in more detail, the very low theoretical gain even with a perfect overlapping discourages further activities in this direction.

### 3.2.1.3 All-to-all

Most of the application time is consumed during collective all-to-all communications. Theses global communications are the key factor for scalability. The impact of all-to-all communication has already been discussed in previous deliverables, notably D1.6 and D1.7, with a specific emphasis on network topologies. In the research field a lot of work has already been done for all-to-all communication with often tradeoff and interconnect specifics optimizations notably switching from inter and intra-nodes with grouped communications. Most of this is already handled by state-of-the-art MPI implementations. However, as demonstrated in the section about GPU optimization of all-to-all communication, a pseudo complete graph of physical interconnects allows reaching peak performance by removing graph contention and benefitting from the full parallel bandwidth available.

#### 3.2.1.4 Disabling MPI Barrier

There is some comment about progress in communication via MPI_Barrier scattered throughout the code. MPI_Barrier can make communication progression but it is not mandatory. By disabling MPI Barrier there is a small performance gain. In addition, from a profiling point of view, measuring timing between each Barrier could create more contention. Even with a low MPI Barrier synchronization cost, the sum of measured timing between them is not necessarily representative of the same run without Barrier. For example, from a software point of view, MPI pre-registered buffers are limited in number and it could lead to another code path in the MPI library if we synchronize processes just before communication, removing asynchronism benefit.

### 3.2.2 Results

For the energy consumption we used both hardware and software tools to provide valid and realistic results. We used a specific electronic device (from ZIMMER) to get the power consumption of a whole chassis containing four bi-sockets nodes running TCo639. It also validated software measures which were used for a larger number of nodes. We also added the theoretical consumption based on the time step multiplied by the number of time steps and average power which is practically the same metric used for GPU. The major difference is that our average power is based on whole chassis measurement presented below (including power supply part).

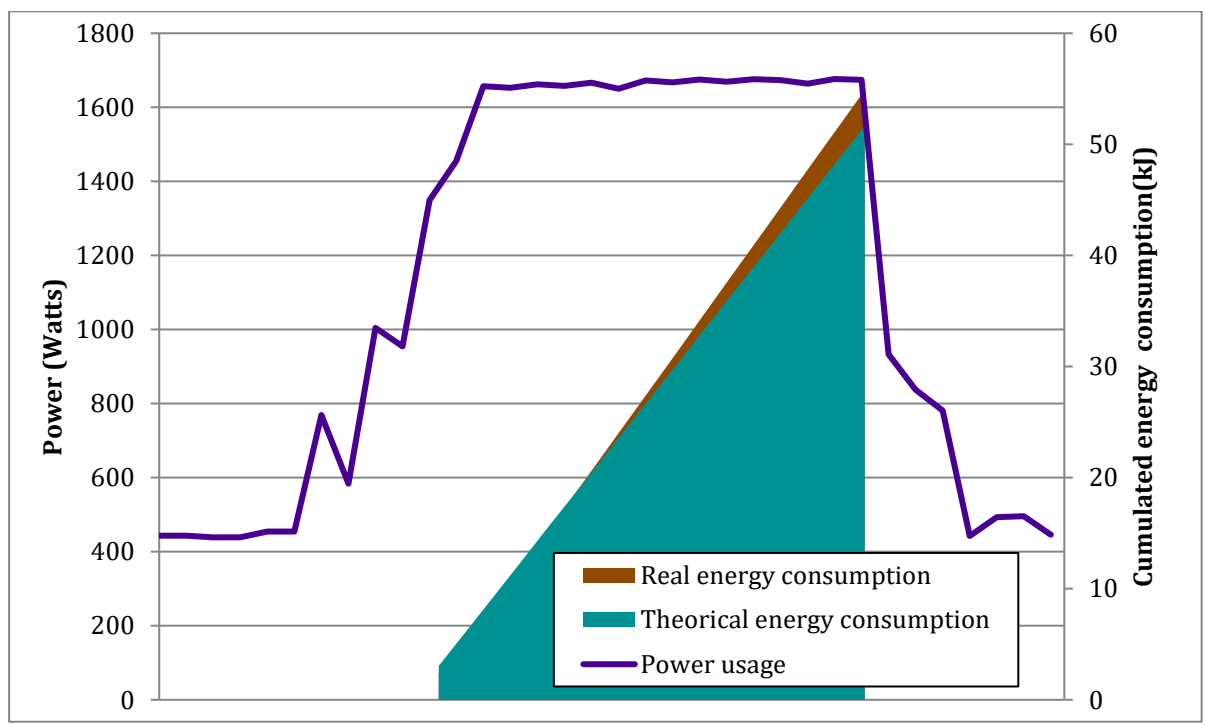

*Figure 10 : Power and energy consumption of one frame (four nodes) during TCo639. Measured with a ZIMMER.*

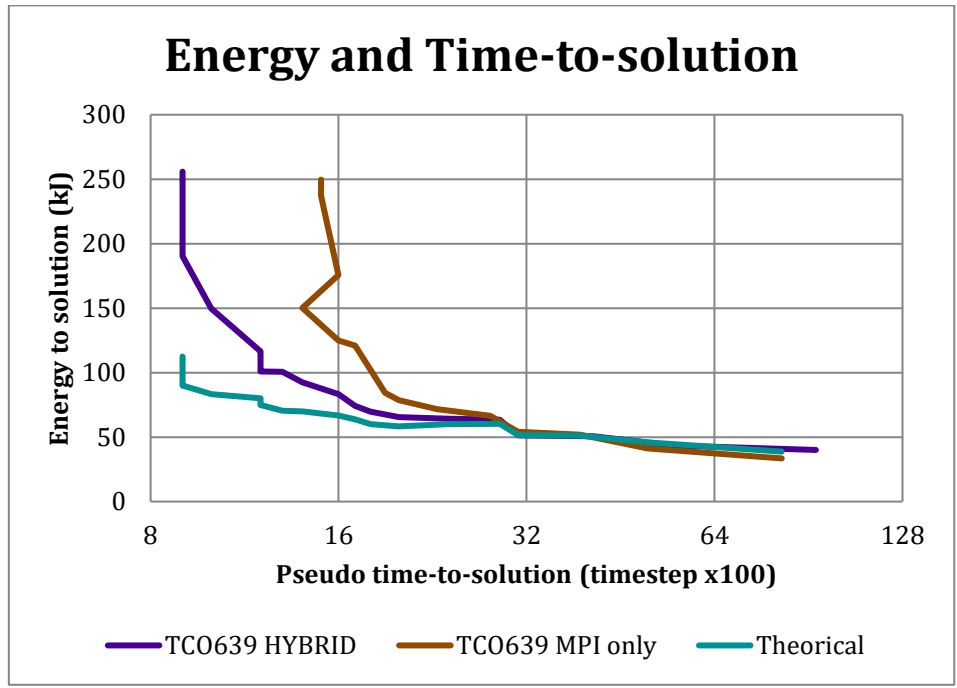

*Figure 11 : Energy consumption is measured via software on each node for a full run.*

Below are presented performances results with current and future production resolutions (Tco1279@9km and Tco1999@5km) and also Tco639@18km resolution. Efficiency and performances are based on time step.

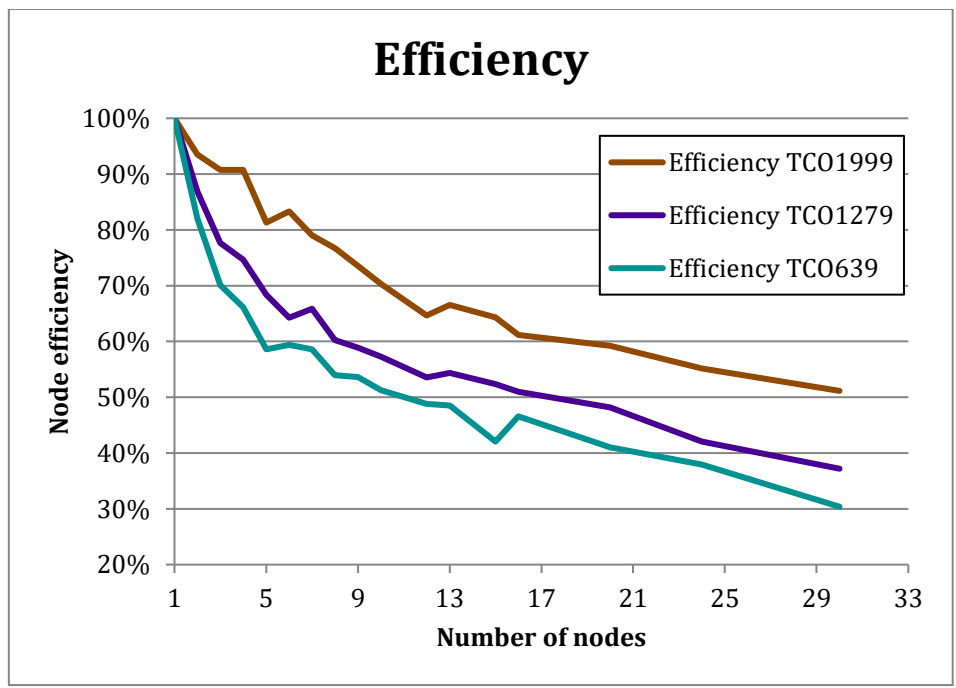

*Figure 12: Efficiency according to the number of nodes for different Tco resolutions.*

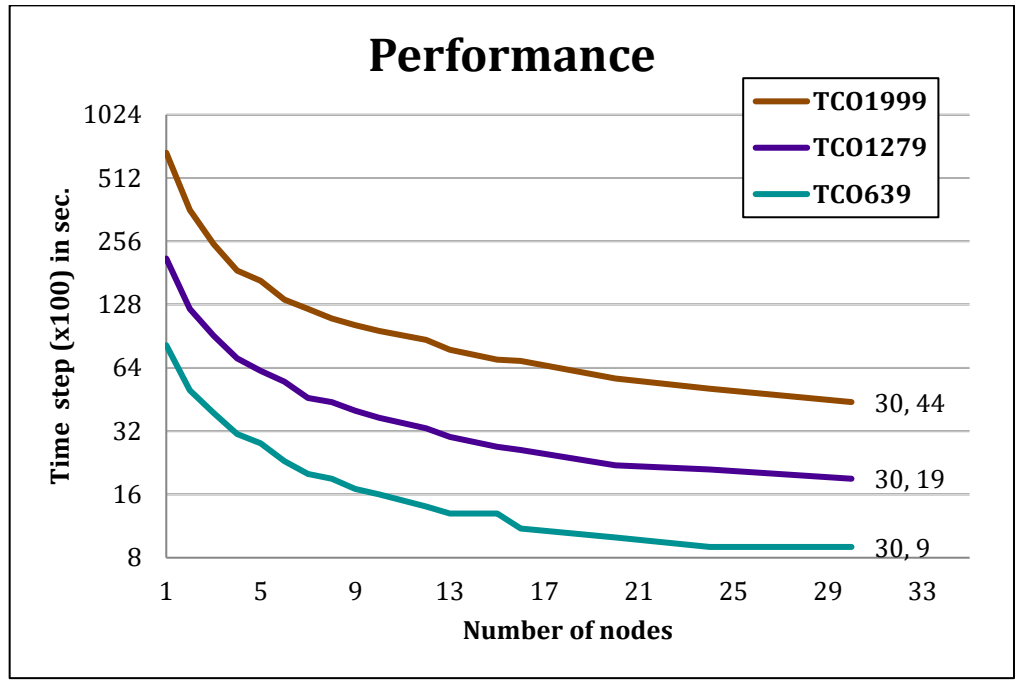

*Figure 13: Performance results according to the number of nodes for different resolution of Tco.*

Below are presented memory usage with different test cases. Hybrid version use threads which only add a new stack inside the MPI process and so reduce memory usage compared to MPI only version. However further work should be done with another measuring tool to confirms and validate our results.

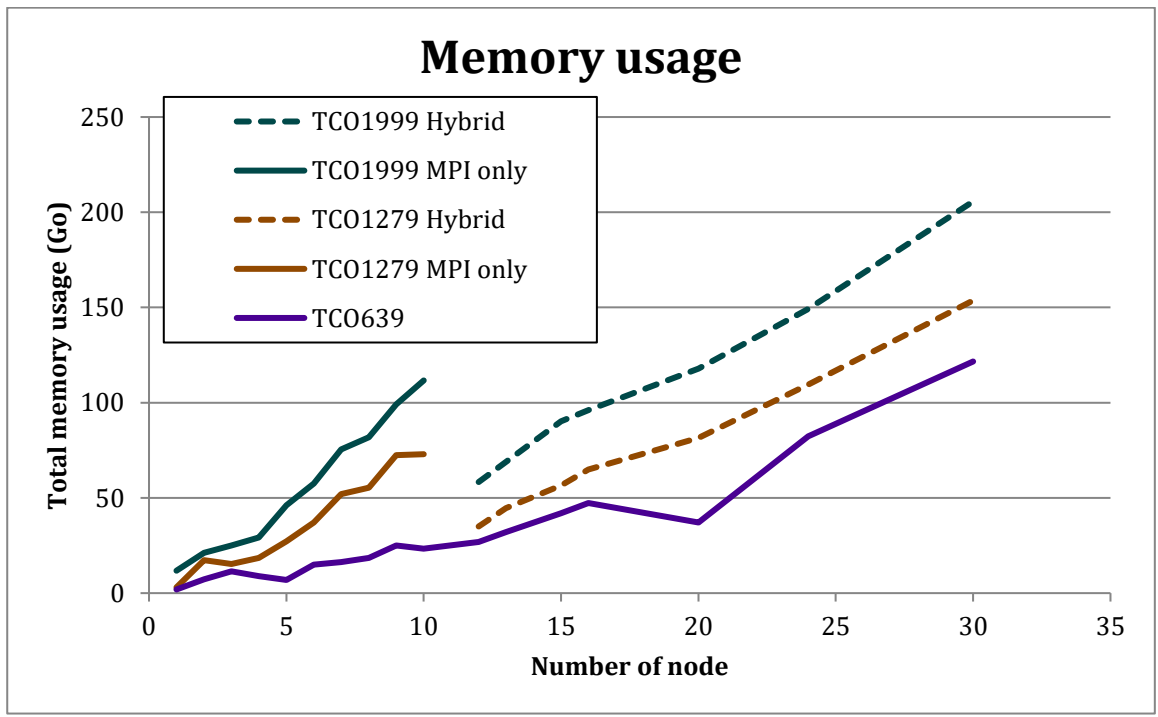

*Figure 14: Memory usage with different test cases.*

### 3.2.3 Conclusion

Because of Intel KNL and KNC being discontinued we choose to focus on Intel Skylake processors which are compatible with the AVX-512 instructions, allowing to reuse the D3.3 KNL optimizations. Our profiling and benchmarks lead us to concentrate on communications. We showed that point-to-point communications are dominated by data preparation and we improved their performance. For the configuration of this specific dwarf, we showed that MPI only in a SMP configuration can be comparable with hybrid OpenMP/MPI notably because of the MPI domain decomposition imbalance and also thanks to the IntelMPI shared memory module. It leads us to optimize at system level with an adapted domain binding.

From theses optimizations we measured performances and memory usage at current and future resolutions (18 km, 9 km and 5 km) and whole energy consumption at 18 km providing realistic production results.

## 4  MPDATA Dwarf

### 4.1  Single-GPU Optimization

#### 4.1.1  Optimization strategies

We optimized all the kernels in MPDATA in a similar fashion to the work described in Deliverable D3.3 (in the MPDATA and Spherical Harmonics sections). A key issue for MPDATA was that in the original implementation there existed a naïve mapping of loops to the GPU, using "kernels" and/or "loop" directives without any code restructuring, and the resulting decomposition chosen by compiler did not work well since runtime loop extents were a poor match to the GPU architecture. In our optimized version we have restructured the code such that such that loops are tightly nested, and we use the "collapse" OpenACC clause to collapse these into a single loop which allows the compiler to map all the inherent parallelism in the algorithm to the GPU parallel architecture in an efficient manner.

We also optimized data management, in a similar fashion to that described for Spherical Harmonics in Deliverable D3.3, such that the fields stay resident on the GPU for the whole timestep loop: all allocations/frees have been moved outside the timestep loop with temporary work arrays being re-used, and all host/device data transfer has been minimized.

#### 4.1.2  Results

The results presented in this section have been obtained using the following software configuration: PGI 17.9, CUDA 8.0.61 and OpenMPI 1.10.7.

In Figure 1 we show the effect of single-GPU optimizations for the MPDATA 512 test case on a single V100 GPU. As described in Deliverable D3.3, the "Roofline" gives a measure of the highest achievable performance that can be obtained on the architecture in question dependent on its available memory bandwidth (as measured using the STREAM benchmark) and peak compute capability. Since MPDATA has a relatively low arithmetic intensity throughout, without exception, it is strictly limited by memory bandwidth. We can see that the optimizations have had the effect of improving performance by a factor of 57X, and that the optimized code is now very close to the roofline. (We also see that the arithmetic intensity has decreased: this is mainly due to the fact that in one kernel there were originally a large number of unnecessary duplicated calculations. Note the removal of these was not a major reason for the overall improved performance.)

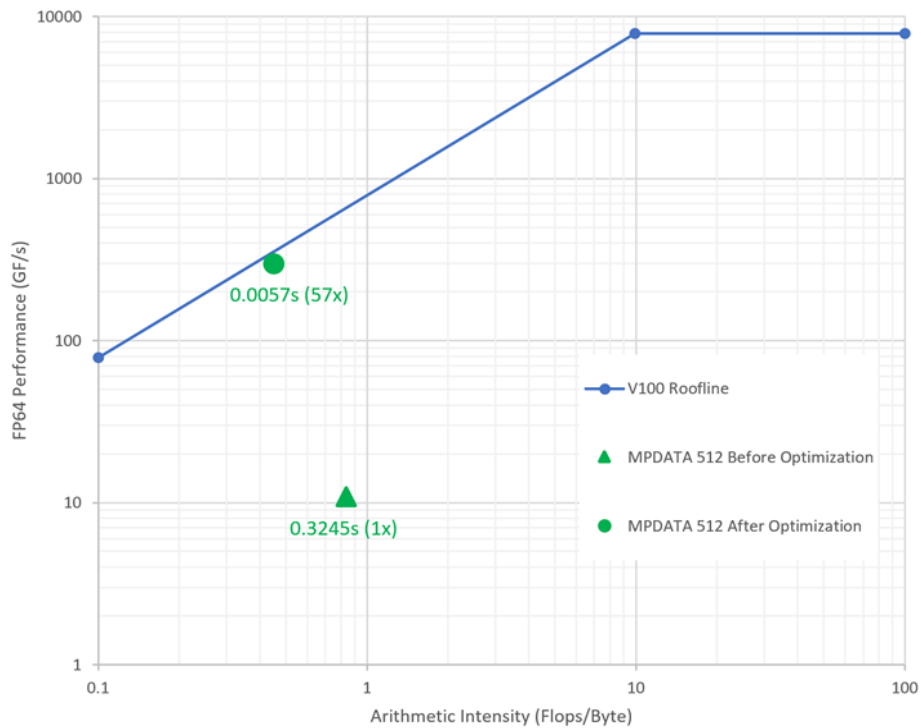

*Figure 15: The performance of the original and optimized versions of MPDATA for the 512 test case on the NVIDIA Tesla V100 GPU, in the context of the roofline for this architecture. The points are positioned in the plot according to their arithmetic intensity: points under the sloping region of the roofline are limited by available memory bandwidth, and points under the horizontal region are limited by peak computational performance.*

In Figure 2 we show the performance of each kernel individually on a bubble plot, where the size of each bubble is proportional to the runtime of that kernel, and the height of the bubble corresponds to the percentage of roofline (i.e. STREAM benchmark memory bandwidth) performance for that kernel. Therefore, the higher the better, and the larger the bubble the more important in terms of overall performance: small bubbles have limited significance and can be ignored. It can be seen that all kernels except one are in the 80-100% region, with most in the upper part of that region. The one exception at ~60% (which has a slightly more complicated structure than the others and needs further investigation) is not a large bubble, so has limited impact on overall performance.

*Figure 16: The performance of each kernel in the MPDATA 512, given as a percentage of the NVIDIA V100 GPU roofline (i.e. STREAM benchmark memory bandwidth since all kernels have a relatively low arithmetic intensity). The size of each bubble is proportional to the time taken by that kernel.*

### 4.2 Multi-GPU Optimization

#### 4.2.1 Introduction

In Deliverable D3.3, we described optimization of a single (computationally demanding) MPDATA kernel and presented performance results on the NVIDIA Tesla P100 (Pascal) architecture. In this deliverable, we substantially extend the work with optimizations and performance results spanning the whole dwarf on a single GPU and on multiple GPUs in parallel. We also take the opportunity to use, for our performance results, the newer V100 (Volta) architecture. Throughout this section, we use V100 GPUs within a DGX-1V server, the architecture of which is described in detail in the Spherical Harmonics section in this deliverable.

In the "Single-GPU Optimization and Results" section we show how single-GPU optimizations can push performance very close to the roofline, and in the "Multi-GPU Optimization and Results" section we describe work undertaken on a custom CUDA-aware MPI implementation of the necessary halo-exchange mechanism to allow utilization of multiple GPUs without the substantial data transfer overheads that occur in the original naïve implementation. Finally, in the "Power and Energy Usage" section we present an analysis of how these aspects vary with the number of GPUs in use.

#### 4.2.2 Optimization strategies

##### 4.2.2.1 Implementation of Custom CUDA-aware MPI Halo Exchange

When utilizing multiple GPUs, two halo exchanges are required each timestep for the zdivVD and pD fields respectively. At the application level, the original mechanism involved code such as:

```
!$ACC update self(zdivVD)
call halo_exchange%execute(zdivVD)
```

```
!$ACC update device(zdivVD)
```

The halo_exchange%execute subroutine is provided by the ATLAS library, which expects the data to be resident on the host CPU. Therefore, the OpenACC directives are required to copy the array from GPU to CPU before the operation, and back to the GPU after. As will be seen, this data movement has huge overhead. Therefore, we developed a new custom halo-exchange mechanism to be used in place of the original, which allows data to be kept resident on the GPU throughout, which we will now describe.

With structured grids, halo exchanges are relatively straightforward since the halo data on each subdomain corresponds to a small subset of "neighboring" subdomains in a clear manner. However, the data structures in this dwarf are unstructured grids, and as a result each subdomain may have halo data elements corresponding to any of the other subdomains. Therefore, the halo exchange requires an "all-to-all" communication pattern, where the sizes vary. We implement this as follows:

- Pack "send" buffer containing "edge" data on each GPU (using OpenACC), for each corresponding remote GPU.

- Exchange this data using CUDA-aware MPI_Alltoallv.

- Unpack "receive" buffer on each GPU (using OpenACC), with each element being stored in its appropriate halo location.

Note that the "Spherical Harmonics" section of this document has information on how to use CUDA-aware MPI with OpenACC.

The above mechanism requires a substantial amount of one-off preparation:

- Use ATLAS API functions to obtain information on which elements in the array are ghost (halo) elements on each GPU.

- Query each halo element (again using the ATLAS API) to obtain information on which remote location it corresponds to.

- Use MPI_Alltoall calls to share all local-remote indexing information amongst all GPUs, such that send/receive buffers can be packed/unpacked correctly.

- Allocate arrays to hold indexing information and send/receive buffers.

This one-off stage is only performed in the first timestep, with the Fortran "save" attribute used to save information for all future timesteps.

### 4.2.3 Results

#### 4.2.3.1 Performance results

We now present results for the MPDATA 1024 test case: larger than the 512 case presented above, the 1024 case is suitable not just for a single GPU but can be decomposed between multiple GPUs in parallel. When only a single GPU is in use, no halo exchange is required, and the timestep time is measured (using the optimizations described above) to be 0.0214s.  But when we enable the halo exchange (still on a single GPU), using the pre-existing communication mechanism, then host/device data transfers occur for the entire field which has a huge overhead, and the time increases by an unworkable factor of 3.9X to 0.0835s. With our new mechanism, the overhead is only around 1% with the time at 0.0216s. By keeping data resident on the GPU we have removed the overhead and allowed effective multi-GPI scaling as we will now show.

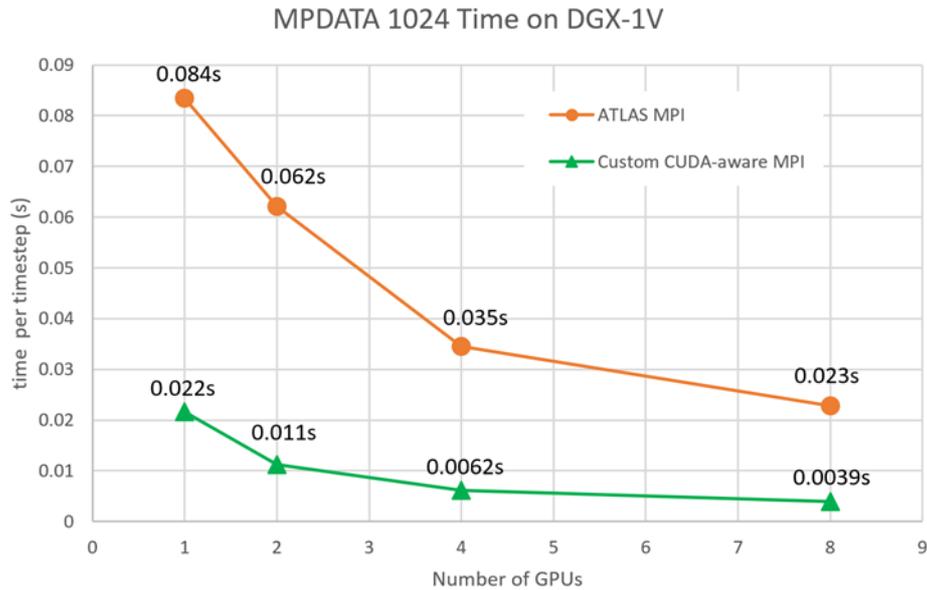

*Figure 17: The dependence of the MPDATA timestep time, for the 1024 test case, on the number of V100 GPUs in use within a single DGX-1V server. Orange circles denote the original halo-exchange communication mechanism which involves host-device data transfer and the host-based ATLAS library. Green triangles denote use of a new custom CUDA-aware halo-exchange mechanism.*

In Figure 7 we show effect of the new custom halo exchange mechanism when running on multiple GPUs, for the MPDATA 1024 test case. Comparing the pre-existing communication mechanism (orange circles) to the new version (green triangles), the reduction timestep time resulting from the removal of data movement overhead is clear, and the code can now scale effectively to the full 8 GPUs available in the DGX-1V server.

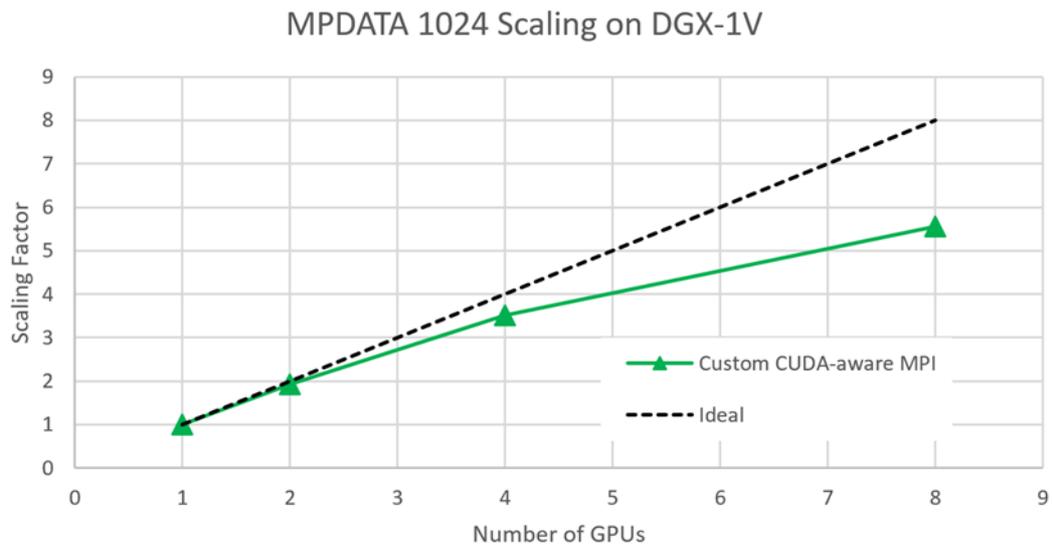

*Figure 18: The scaling of the MPDATA, for the 1024 test case, with increasing numbers of V100 GPUs within a single DGX-1V server, where the new custom CUDA-aware halo-exchange mechanism is in use. Also shown is a line denoting ideal scaling.*

In Figure 4 we show the scaling factor (the time taken on 1 GPU divided by the time taken on the specific number of GPUs in use), using the new mechanism, and compare with the ideal scaling line. In the "Sperical Harmonics Dwarf" section of this deliverable, we describe the DGX-1V architecture and discuss its hardware characteristics in relation to all-to-all communications. Since we are also performing all-to-all communications in this case, then the same issues are relevant, but to a lesser extent since in this case the global data volume involved in each all-to-all is only a fraction of the full field volume (that corresponding to the halo region), and not the full field volume (which is the case for Spherical Harmonics). So, the scaling is better in this case, but we still clearly see a "kink" in the line when going from 4 to 8 GPUs resulting from the fact that the server has full-NVLink-connectivity when using 4 GPUs but not 8 (in which case some communications must go through PCIe and QPI). The Spherical Harmonics results also strongly suggest that the scaling for MPDATA would be substantially better when using the newly-announced high-bandwidth NVSwitch interconnect, and we will perform such benchmarking in future work on the upcoming DGX-2 server which features 16 GPUs interconnected with NVSwitch in an all-to-all topology.

#### 4.2.3.2 Power and Energy Usage

We used the nvidia-smi tool to measure GPU power draw during the MPDATA timestep loop (for the 1024 test case), in a similar way to that described for the Spherical Harmonics Dwarf in this deliverable, using our optimized code. We found that the average draw for a single GPU was 196W, and this decreased with the number of GPUs to 183W, 173W and 152W for 2, 4, and 8 GPUs respectively. The explanation is: the further the deviation from ideal scaling, the more each GPU is spending idle and the lower the draw. As discussed in the Spherical Harmonics section, these power numbers only account for the GPUs in the node but they should still give good indications since the GPUs will dominate the power draw of the node.

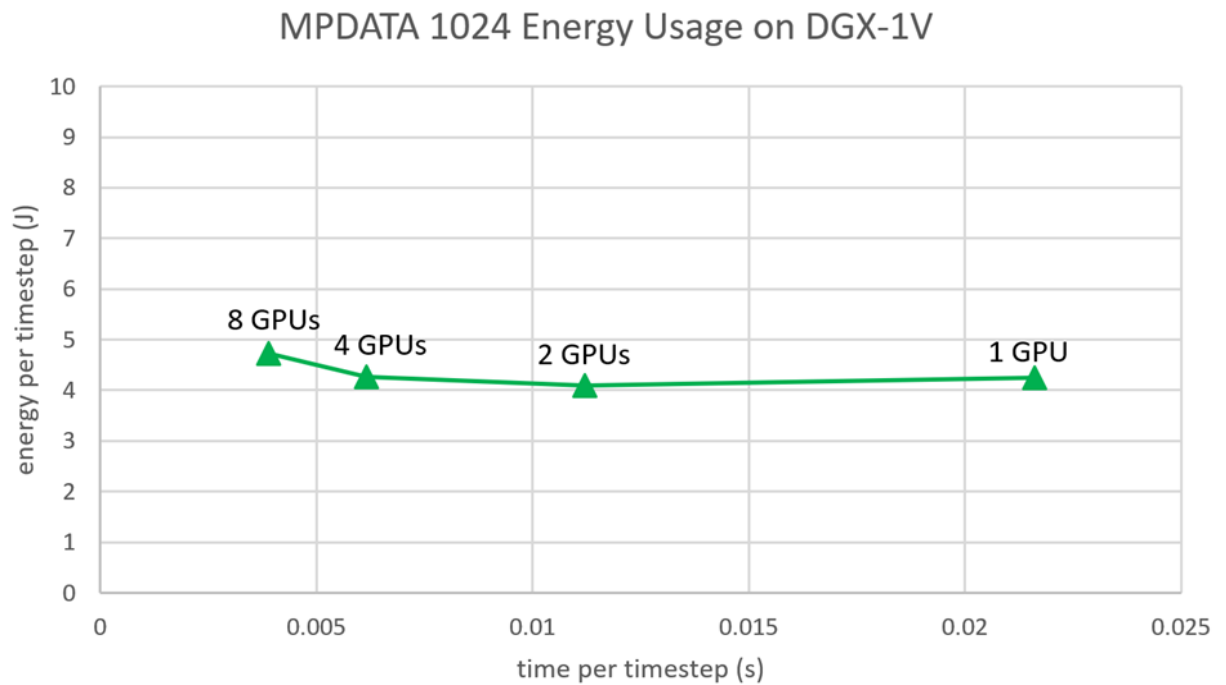

*Figure 19: The Energy usage of the MPDATA 1024 test case as a function of the time taken, per timestep. The different points correspond to different numbers of GPUs in use, as shown.*

In Figure 5 we can combine these power draw figures with the timestep times (given in Figure 3) to show how the energy used per timestep varies with the number of GPUs (and in turn the timestep time, which is the variable used on the x-axis). The series is observed to be fairly flat, which shows that the energy usage is largely independent of the number of GPUs, with the exception that there is a modest increase when using all 8 GPUs to reach the lowest timestep time. All energy values are within the range of 4J to 5J per timestep, however. In other words, as the number of GPUs increases, the reduction in power draw described above largely compensates for the increase in overall time (across all GPUs - i.e. the deviation from ideal scaling).

### 4.2.4 Summary

We showed that, with optimizations designed to expose parallelism as efficiently as possible and minimize data management overheads, the single-GPU performance of the MPDATA dwarf is now very close to optimal performance as given by the roofline for the GPU architecture in use (i.e. the measured STREAM benchmark bandwidth since the dwarf is completely memory bandwidth bound).

We described development of a new custom implementation of the halo exchange mechanism, required when multiple GPUs are being utilized in parallel, which uses CUDA-aware MPI to avoid the huge data movement overheads associated with the original version and we demonstrated scaling to 8 GPUs within a DGX-1V server. While scaling is observed to the full 8 GPUs, there is deviation from ideal scaling and we believe that performance will be improved on the newly announced DGX-2 server which features the high-bandwidth NVSwitch interconnect with all-to-all topology – this will be tested in future work.

We also showed, through power draw measurements, that the energy used per timestep is largely independent of the number of GPUs being utilized within the server, with only slight overheads at the higher GPU counts.

### 4.3 CPU based Multi-node Optimization

#### 4.3.1 Profiling

The MPDATA dwarf has been profiled with Intel APS. The test case is the diagonal configuration on the grid O1024. The runs have been launched on 2 nodes, 64 nodes and 128 nodes, all with 32 OpenMP threads (1 MPI task per node).

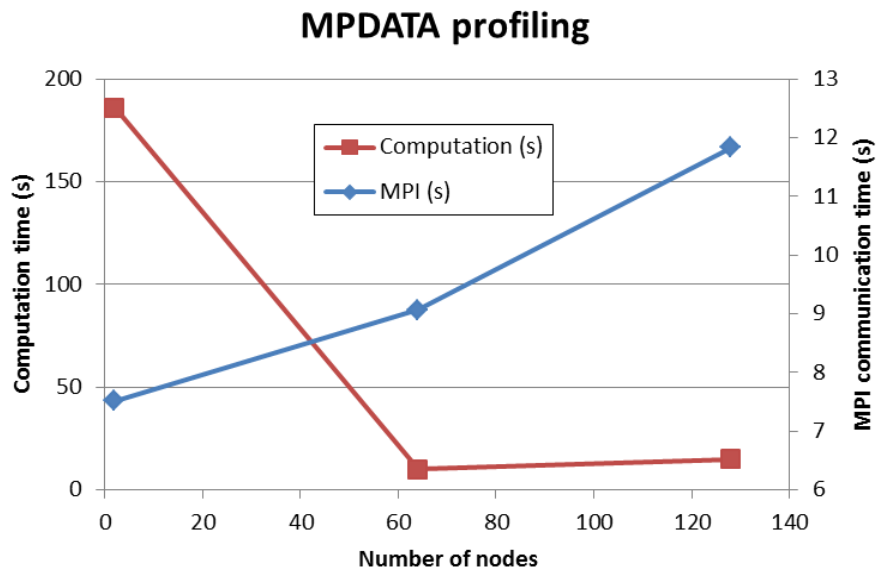

*Figure 20: MPDATA profiling results showing the time spent in computation (left vertical axe) and in MPI communications (right vertical axe) on O1024 test case.*

As one can see in Figure 20 the MPI communication time is growing with the number of nodes. The dwarf is MPI bound with 47% and 44% of the elapsed time with 64 and 128 nodes respectively.

It should be noted that the test case does not scale well on 128 nodes as compared with the run on 64 nodes. There are some scalability issues, maybe due to a few number of grid points per cores.

#### 4.3.2 Optimization strategies

The MPI library, especially the Intel MPI one, relies on several interconnection protocols. It has to be wisely chosen according to the cluster hardware. In the benchmark cluster the interconnect network is based on Infiniband protocol. Traditional Infiniband support uses the Reliable Connection (RC) protocol to exchange MPI messages, but the User Datagram (UD) protocol has emerged as a lower memory consumption, more scalable alternative. The first optimization has been to enable this protocol in the Intel MPI library.

A second optimization is the replacement of the "manual" implementation of the AlltoAllV algorithm inside the halo_exchange subroutine by the MPI_AlToAllV function.

To enhance the scalability, one may consider overlapping the MPI communications with computation loops. This has been introduced by two variants based on a single core idea: the computation loops have to be evaluated on the data to be exchanged before sending them and then they are evaluated on the private data while the communications are received. The first one is implemented via a mask array separating the shared versus the private data. The second one is implemented by an indirection array storing the shared (resp. the private) indices of the data.

### 4.3.3 Results

The modified versions have been launched on a cluster with nodes based on Skylake 2630 processors (16 cores, 192GB) interconnected by an Infiniband EDR network.

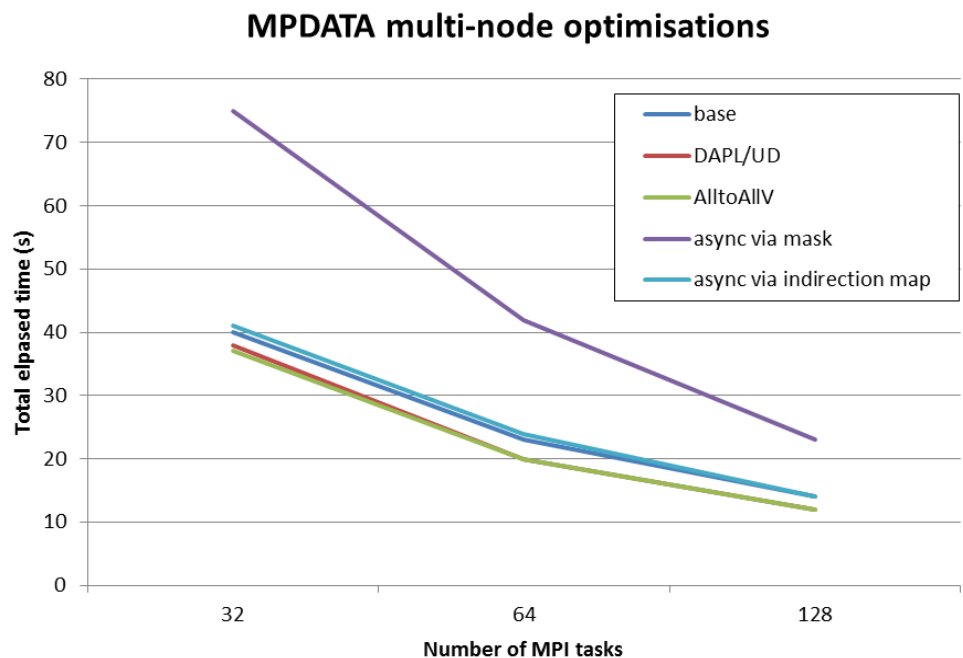

Figure 21: MPDATA scalability results of multi-node optimizations on a 16 physical cores SKX with 1 MPI task per socket and 1 OpenMP thread per physical core

Figure 12 shows scalability results of multi-node optimizations on a 16 physical cores SKX with 1 MPI task per socket and 1 OpenMP thread per physical core.

The two combined optimizations (UD Infiniband protocol and AllToAllV) show a speedup of nearly 15% (20s vs 23s on 64cores, 12s vs 14s on 128 cores).

The two variants of the async patterns are less performant than the AlltoAllv implementation. In the first variant, the loops are executed two times with a conditional inside them. This alters the performance. And the second one, the data are not contiguous anymore. A further implementation could be a renumbering of the data in order to split the private and the shared sets in two disjoint contiguous sets.

Figure 22 shows the energy versus the time to solution. This has been obtained considering the original dwarf implementation with the first two optimizations (aka the optimized protocol and the alltoallv from the MPI library).

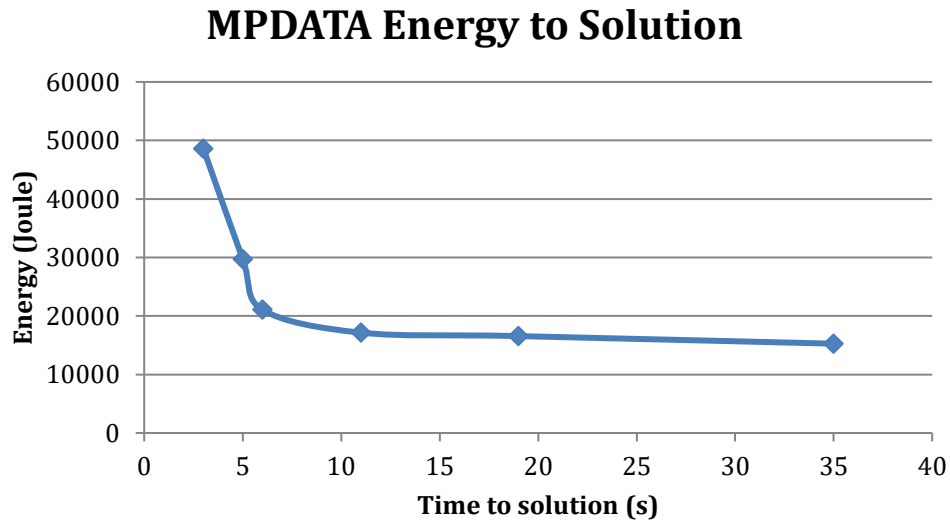

*Figure 22: MPDATA energy to solution results of multi-node optimizations on a 16 physical cores SKX with 1 MPI task per socket and 1 OpenMP thread per physical core.*

#### 4.3.4 Conclusion

We showed that the MPDATA dwarf is communication bound on a CPU system in the strong scaling regime. Optimizing the usage of Intel MPI library shows some improvements. A faster protocol has been selected. And the custom hand written AlltoAll has been replaced by the MPI library version of the function MPI_AllToAllV. Combined these two optimizations gives roughly 15% of speed-up.

To enhance the scalability, two variants have been implemented in the idea of overlapping the MPI communication with computation loops. The computational loops have therefore been modified. The whole elapsed time is at best equal to the elapsed time of unmodified version. A further insight could be gained by profiling both the MPI overlapping and the loop vectorization.

## 5  Conclusion

In this task, the Spectral Transform – Spherical Harmonics and MPDATA dwarfs have been optimized and benchmarked for both multi-GPU and CPU multi-node configurations. Different kinds of approaches were considered regarding the specificities of each hardware configuration. On a DGX-1 and DGX-2, the hardware provides a complete graph of interconnection which allows shared memory specifics optimizations and performances get close to the GPU roofline. We have described custom implementations of data exchanges, which uses CUDA-aware MPI to avoid data movement between CPU and GPU and optimized the NVLINK bandwidth usage. We presented the energy consumptions results of 4 V100 GPUs within a DGX-1 server.

For CPU based multi-nodes we focus on optimizing inter-node communication. We detailed point-to-point and all-to-all optimizations. We also gave global energy consumption measurements in addition with performances and memory usage at different resolutions (up-to 5km). As a conclusion the different optimizations approaches and results presented in D3.3 and the current report (D3.4) lead to a better understanding of theses dwarf boundaries and will be profitable for the NWP community. Indeed lessons learned from these experiments, targeting multi-GPU and CPU based multi-node, could benefit from each other in a complementary way. As an example, improving data locality as presented for GPU systems can be applied for CPU also, at the same time, improving communications between nodes can be applied on a multi-node GPU-accelerated system. Moreover, an hybrid approach can be considered.

## Document History

| Version | Author(s) | Date | Changes |
|---|---|---|---|
| 0.1 | BULL (ER) | 16/04/2018 | Initial version. |
| 0.2 | BULL (ER) | 26/04/2018 | Initial report plan with draft intro. |
| 0.3 | BULL (ER, DG, LD), NVIDIA (AG) | 18/05/2018 | Integration of both GPU and CPU. |
| 0.4 | BULL (ER, DG, LD), NVIDIA (PM) | 24/05/2018 | Rewording and cosmetic changes. |
| 1.0 | BULL (ER, DG, LD), NVIDIA (AG) | 30/05/2018 | Update according to internal reviews. |

## Internal Review History

| Internal Reviewers | Date | Comments |
|---|---|---|
| Oliver Fuhrer (MSwiss) | 26/05/2018 | Approved with comments |
| Peter Bauer (ECMWF) | 26/05/2018 | Approved with comments |
|  |  |  |
|  |  |  |

## Effort Contributions per Partner

| Partner | Efforts |
|---|---|
| BULL | 8.04 PM |
| NVIDIA | 5.61 PM |
| **Total** | **13.65 PM** |

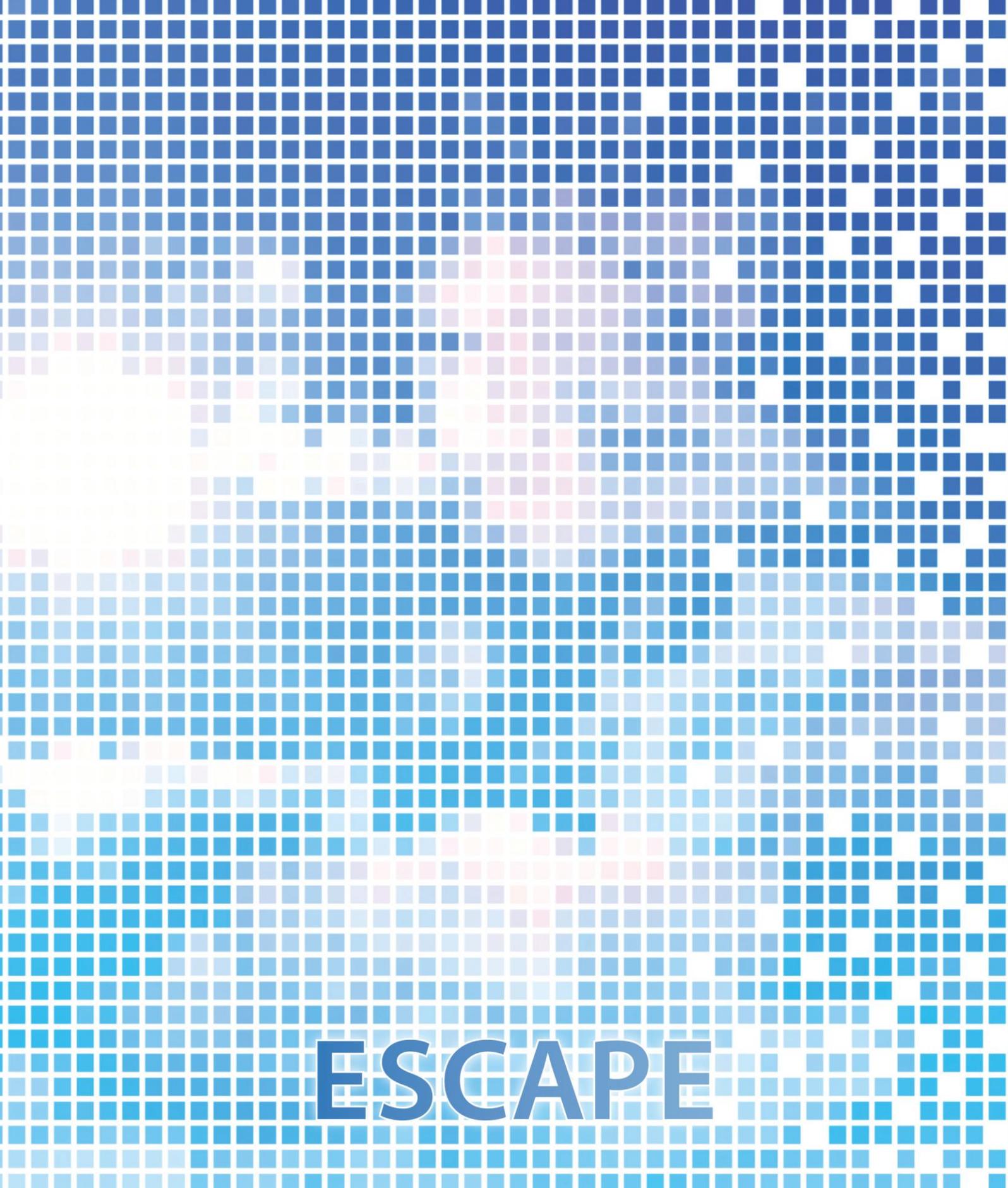